\documentclass[fleqn,usenatbib,useAMS]{mnras}

\usepackage{graphicx}	
\usepackage{amsmath}
\usepackage{amssymb}	
\usepackage{bm}
\usepackage{physics} 
\usepackage{multicol}

\newcommand{\OmK}{\Omega_\text{K}}

\defcitealias{cb20}{CB20}
\defcitealias{cb21}{CB21}

\title[]{Turbulence in outer protoplanetary disks: MRI or VSI?}
\author[]{Can Cui$^{1}$\thanks{E-mail: \href{mailto:cc795@cam.ac.uk}{cc795@cam.ac.uk}} 
and Xue-Ning Bai$^{2,3}$\thanks{E-mail: \href{mailto:xbai@mail.tsinghua.edu.cn}{xbai@mail.tsinghua.edu.cn}} \\
$^{1}$DAMTP, University of Cambridge, CMS, Wilberforce Road, Cambridge CB3 0WA, UK \\
$^{2}$Institute for Advanced Study, Tsinghua University, Beĳing 100084, China  \\
$^{3}$Department of Astronomy, Tsinghua University, Beĳing 100084, China}

\pubyear{2021}

\begin{document}
\label{firstpage}
\pagerange{\pageref{firstpage}--\pageref{lastpage}}
\maketitle

\begin{abstract}

 The outer protoplanetary disks (PPDs) can be subject to the magnetorotational instability (MRI) and the vertical shear instability (VSI). While both processes can drive turbulence in the disk, existing numerical simulations have studied them separately. In this paper, we conduct global 3D non-ideal magnetohydrodynamic (MHD) simulations for outer PPDs with ambipolar diffusion and instantaneous cooling, and hence conductive to both instabilities. Given the range of ambipolar Els\"{a}sser numbers ($Am$) explored, it is found that the VSI turbulence dominates over the MRI when ambipolar diffusion is strong ($Am=0.1$); the VSI and MRI can co-exist for $Am=1$; and the VSI is overwhelmed by the MRI when ambipolar diffusion is weak ($Am=10$). Angular momentum transport process is primarily driven by MHD winds, while viscous accretion due to MRI and/or VSI turbulence makes a moderate contribution in most cases. Spontaneous magnetic flux concentration and formation of annular substructures remain robust in strong ambipolar diffusion dominated disks ($Am\leq1$) with the presence of the VSI. Ambipolar diffusion is the major contributor to the magnetic flux concentration phenomenon rather than advection.
\end{abstract}

\begin{keywords}
instabilities -- MHD -- turbulence  -- methods: numerical -- protoplanetary discs
\end{keywords}

%%%%%%%%%%%%%%%%%%%%%%%%%%%%%%%%%%%%%%%%%%%%%%%%%%
\section{Introduction}\label{in}

Turbulence is essential to many fundamental processes in protoplanetary disks (PPDs), including angular momentum transport, dust dynamics, planet migration, and chemical mixing. Turbulence ensues as a consequence of hydrodynamic or magnetohydrodynamic (MHD) instabilities. Seeking observational evidence of these instabilities and distinguishing one from another have none the less been challenging. The unprecedented sensitivity and resolution of Atacama Large Millimeter/submillimeter Array (ALMA) brings a step closer to reveal the origins and nature of the turbulence in PPDs, being capable of probing turbulence strengths \citep{teague+16,flaherty_etal17,flaherty_etal18,flaherty_etal20} and gas velocity structures \citep{teague+19,teague+21} in the disk.

The magnetorotational instability (MRI; \citealp{bh91}) was considered to be the primary source that drives turbulence and angular momentum transport in PPDs. However, the extremely weak level of ionization inevitably induces non-ideal MHD effects, which can weaken or suppress the MRI. Instead, magnetized disk winds are found to dominate the disk accretion \citep[e.g.,][]{bs13,bai17,bethune_etal17,wang_etal19,gressel_etal20}. In outer regions of the disk ($\gtrsim30$ AU), numerical simulations demonstrated that the MRI is dampened due to ambipolar diffusion, yielding weak levels of turbulence \citep[e.g.,][]{simon_etal13a,simon_etal13b,lesur_etal14,bai15,simon_etal18}. Latest global 3D non-ideal MHD simulations confirmed this result and further illustrated the coexistence of the weak MRI turbulence and MHD winds \citep[][hereafter \citetalias{cb21}]{cb21}.

As alternative sources of turbulence, purely hydrodynamic mechanisms have caught attention \citep[][]{lu19,lesur+22}. Among them, the vertical shear instability (VSI) is the disk analogue of the Goldreich--Schubert--Fricke instability \citep{gs67}. Because the stratification and thermodynamic conditions are very conductive to its onset, the VSI may span a large extent, likely covering the entire outer disks \citep[e.g.,][]{pk19}. Its robustness in the non-ideal MHD dominated regions (\citealp[][hereafter \citetalias{cb20}]{lp18,cb20}; \citealp{cl21,LK22}) and the potential observational significance \citep{lin19,flock_etal20,schafer+20} render the VSI a very promising hydrodynamic mechanism in PPDs.

The MRI and the VSI could both emerge and sustain in outer disks. However, extant published numerical simulations on them have been performed separately. In this work, we explore the interplay between them. We extend the work of \citetalias{cb21}, where they focused on the properties of the MRI by using a cooling timescale comparable to the disk dynamical timescale ($\tau=1$), and hence inhibits the VSI. We also extend the work of \citetalias{cb20}, where they performed simulations on the VSI by 2D non-ideal MHD simulations, and hence precludes the MRI. A three-dimensional, ambipolar diffusion dominated outer disk with instantaneous cooling ($\tau=0$), conductive to the VSI and the MRI, is now explored. 

In addition, \citetalias{cb21} illustrated the presence of magnetic flux concentration, and the subsequent gas annular substructure formation \citep[also known as ``zonal flows",][]{johansen_etal09}, in ambipolar diffusion dominated outer disks. The magnetic flux concentration phenomenon is proved to be robust; it occurs in situations as diverse as local and global, ideal and non-ideal MHD simulations \citep[e.g.,][]{bs14,suriano_etal18,suriano_etal19,rl18,rl19,rl20,JL21,cb21}. Local and global simulations illustrate that the magnetic flux concentration could result in dusty rings and gaps formation \citep{rl20,xb21,hu+22}.
In this work, we further seek to answer that whether the spontaneous magnetic flux concentration phenomenon could exist in VSI turbulent disks.

This paper is organized as follows. In \S\ref{sec:method}, we briefly describe the numerical methods and simulation setup. In \S\ref{sec:re}, we show the simulation results and discuss turbulence properties, angular momentum transport, and mangeitc flux concentraton and annular substructure formation. We finish in \S\ref{sec:cd} with a summary and discussion.

\section{Methods}\label{sec:method}

The numerical setup of this work closely resembles that of \citetalias{cb21} (see their section 2). The instantaneous cooling is now applied to trigger the VSI. Key elements are outlined here.

\subsection{Disk Model}\label{sec:dm}

The disk model is established such that radial temperature and density profiles are set to power laws by $\rho\propto r^{-q_D}$ and $T\propto r^{-q_T}$, where the power-law indices are $q_D=-2, q_T=-1$, respectively \citep[][\citetalias{cb21}]{bs17}.
The gas scale height is $H=c_s/\OmK$, where $c_s^2=P/\rho$ is the squared isothermal sound speed. The disk aspect ratio $H/r$ is a constant given $q_T=-1$. It is set to $H/r=0.1$ within $\pm3.5H$ across the midplane and transitions to $H/r=0.5$ in the atmosphere. The temperature is relaxed to initial equilibrium values over a timescale $\tau_{\rm cool}$. After each simulation time step $\Delta t$, the gas temperature is adjusted by the amount 
\begin{equation}
    \Delta T =(T_\mathrm{eq}-T)[1-\exp(-\frac{\Delta t}{\tau_{\rm cool}})] \ .
\end{equation}
A dimensionless thermal relaxation time is defined as $\tau\equiv\tau_{\rm cool}/P_{\rm orb}$, where local orbital time $P_{\rm orb} (R)= 2\pi/\OmK$, with Keplerian frequency $\OmK=R^{-3/2}$. To trigger the VSI, we adopt a locally isothermal disk with instantaneous cooling $\tau= 0$. 

The initial magnetic fields are purely poloidal by prescribing an azimuthal vector potential. This yields a purely vertical field at the midplane, and its strength is set according to the midplane plasma beta parameter $\beta_0$. Ambipolar diffusion is parameterized by ambipolar Els\"{a}sser number $Am\equiv v_\textrm{A}^2/(\eta_\textrm{A}\Omega_\textrm{K}$), where $v_\textrm{A}$ is the Alfv\'{e}n speed and $\eta_\textrm{A}$ is the ambipolar diffusivity. 
It is taken to be constant within $\pm3.5H$ across the midplane, and gradually transitions to $Am=100$ from the disk towards the atmosphere. 
%$Am\equiv\gamma\rho_i/\Omega_K$, where $\gamma\rho_i$ is the neutral-ion collision frequency. 

\subsection{Numerical Setup}\label{sec:ns}

We use the grid-based high-order Godunov MHD code Athena++ \citep{stone_etal20} to conduct global 3D non-ideal MHD simulations.
We solve the equations of non-ideal MHD using the van Leer time integrator, the HLLD Riemann solver, and the piecewise linear reconstruction. Super time-stepping is employed to speed up calculations for diffusive non-ideal MHD physics. 

The simulations are performed in spherical polar coordinates $(r,\theta,\phi)$. To better present simulation results, cylindrical coordinate system is used with $R=r\sin\theta$ and $z=r\cos\theta$. In code units, $GM=R_0=1$, with $M$ the central star mass and $R_0$ a reference radius at the inner radial boundary. 
The simulation domain spans over $r\in[1,100]$, $\theta\in[0,\pi]$, and $\phi\in[0,\pi/4]$. Three levels of mesh refinement are used as in \citetalias{cb21}, where the most refined disk zone achieves 32 cells$/H$ in $r$, 30 cells$/H$ in $\theta$ at the midplane, and 12 cells$/H$ in $\phi$. 

At the inner radial boundary, we set temperature to their equilibrium values. Density is extrapolated from the last active zone, assuming $\rho\propto r^{-q_D}$. The angular velocity is set to the minimum value between equilibrium $v_\phi$ and the angular velocity of solid body rotation $\Omega_\mathrm{K}(R_0)R$. The other two velocity components are $v_r=v_\theta=0$. 

Other boundary conditions are standard. At outer radial boundary, the hydrodynamic variables are extrapolated from the last active zone, assuming $\rho \propto r^{-q_D}$, $T \propto r^{-q_T}$ and $v_\phi \propto r^{-1/2}$. Radial and meridional velocities $v_r$ and $v_\theta$ are copied directly from the last active zone, except setting $v_r = 0$ when $v_r<0$. Magnetic field variables of the inner and outer ghost cells are set by $B_r \propto r^{-2}$, $B_\theta \propto \mathrm{const.}$, $B_\phi \propto r^{-1}$. 
Our $\theta$ domain reaches to the poles, and polar wedge boundary condition is adopted there \citep{zhu18}. Lastly, our $\phi$ boundary condition is periodic.

\subsection{Simulation Models}\label{sec:sm}

Table \ref{table:runs} lists the simulation models. Model name with a `c' denotes instantaneously cooling in the disk, which are newly conducted. The rest of the simulation models are taken from \citetalias{cb21}. Model name with an `E' indicates those which $E_\phi=0$ is enforced at the inner boundary, in order to stabilize the disk innermost region \citepalias{cb21}.
The midplane disk magnetization of the initial poloidal field is set to $\beta_0=10^4$, where $\beta_0$ is the ratio of gas pressure to magnetic pressure. For typical disk models (Minimum-mass solar nebula \citet{weidenschilling77} or similar models), this level of magnetization yields accretion rates in agreement with observational constraints \citep{simon_etal13b}. We employ three levels of ambipolar diffusion strengths, ${\rm Am}\in\{0.1, 1, 10\}$. In addition, all simulations are run up to $3000P_0$, where $P_0= P_{\rm orb} (R_0)=2\pi$ is the orbital period at the inner boundary.

\begin{table}
 \caption{List of simulation models and parameters. } \label{tab:anysymbols}
 \begin{tabular*}{1\columnwidth}{@{}l@{\hspace*{33pt}}c@{\hspace*{33pt}}c@{\hspace*{33pt}}c@{\hspace*{33pt}}c@{}}
  \hline
  Model & $\beta_0$ & Am & $\tau$ & Runtime$\;(P_0)$ \\
  \hline
  \texttt{Am0.1c} & $10^4 $ & 0.1 & 0 & 3000 \\
  \texttt{Am1c} & $10^4$  & 1 & 0 & 3000 \\
  \texttt{Am10Ec} & $10^4$  &  10 & 0 & 3000 \\
  \hline 
  \texttt{Am0.1} & $10^4 $ & 0.1 & 1 & 3000 \\
  \texttt{Am1} & $10^4$  & 1 & 1 & 3000 \\
  \texttt{Am10E} & $10^4$  &  10 & 1 & 3000 \\
  \hline  
 \end{tabular*}
\label{table:runs} 
\end{table}

\section{Results}\label{sec:re}

In this section, the analyses will mostly focus on the turbulence properties and the magnetic flux concentration of simulations with instantaneous cooling $\tau=0$. The comparison between $\tau=0$ simulations (this work) and $\tau=1$ simulations \citepalias{cb21} will also be covered.
In the interest of brevity, we point the reader to \citetalias{cb21} for more detailed technical descriptions, analyses, and discussions on ambipolar diffusion dominated 3D global simulations. 

\begin{figure*}
\centering
\includegraphics[width=0.95\textwidth]{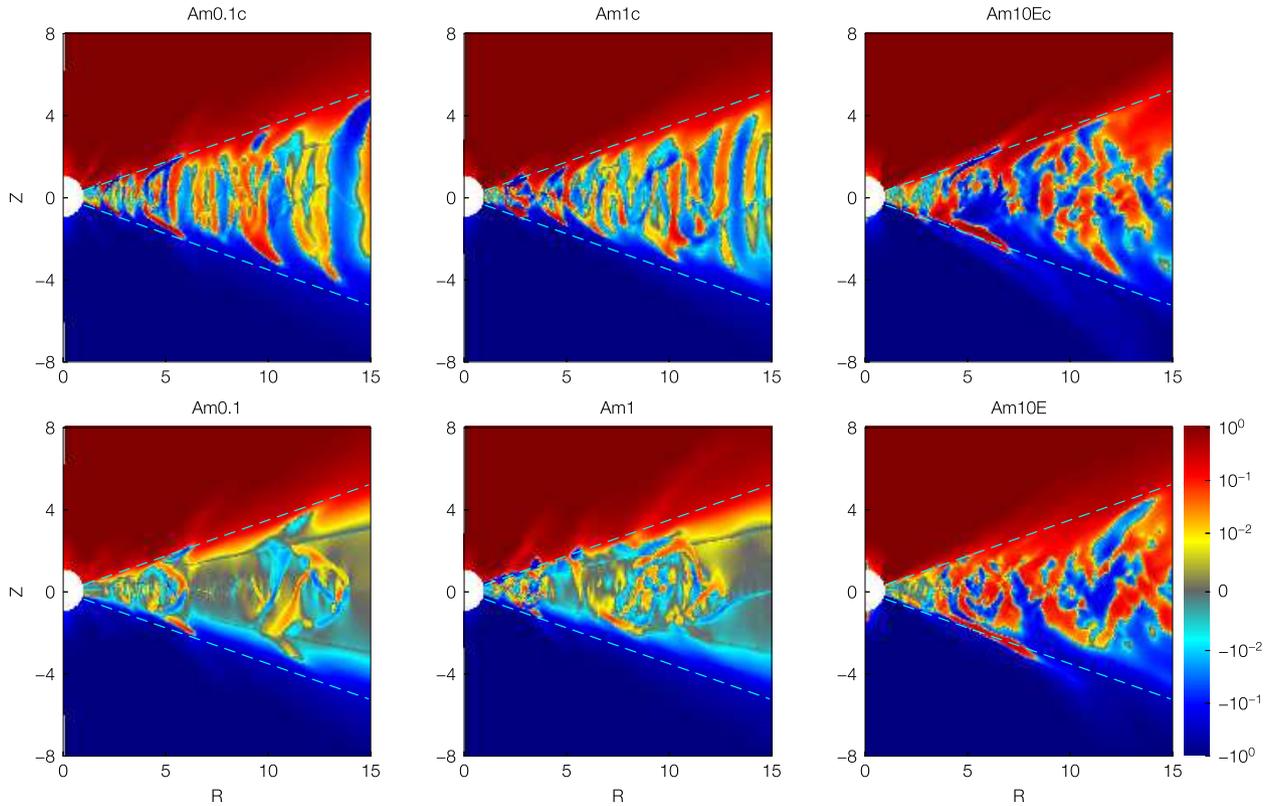}
\caption{Snapshots of azimuthally averaged $v_z/c_s$ for different simulation models at $t = 2600P_0$. Top panels are models with instantaneous cooling $\tau=0$. Bottom panels are models with $\tau=1$. Dashed lines denote $\pm3.5H$ above and below the midplane. They mark the transition from non-ideal MHD dominated disk zone to ideal MHD dominated wind zone.}
\label{fig:vz}
\end{figure*}

\subsection{Overview of gas and magnetic field evolution}\label{sec:3.1}

The evolution of gas and magnetic fields in the disk and wind zone is in a similar manner as of \citetalias{cb21}, except that now in the stronger ambipolar diffusion models (\texttt{Am0.1c} and \texttt{Am1c}), the VSI turbulence sets in. We summarize the gas and magnetic fields evolution as follows.

In the disk zone, where ambipolar diffusion dominates, the gas evolution is determined by the MRI or VSI, or a combination of both, depending on the ambipolar diffusion strength (sections \ref{sec:3.2} and \ref{sec:3.3}). Both types of turbulence contribute to the angular momentum transport process, while MHD winds is the primary source to drive disk accretion (section \ref{sec:3.5}). In the azimuthal plane, vortices are not observed in all of our models (section \ref{sec:3.6}). 

The initial condition of our simulations sets large-scale purely poloidal magnetic fields threading the disk. Poloidal fields are then wound up by Keplerian rotation to generate toroidal fields, which then dominate the magnetic field strength (Figure \ref{fig:B}). In the disk atmosphere, where ideal MHD dominates, the large-scale poloidal fields result in the launching of magnetized winds. In the case of strong ambipolar diffusion models, poloidal magnetic flux accumulates into flux sheets at different radii, yielding annular substructures in radial surface density profiles (section \ref{sec:3.4}).

\begin{figure*}
\centering
\includegraphics[width=0.95\textwidth]{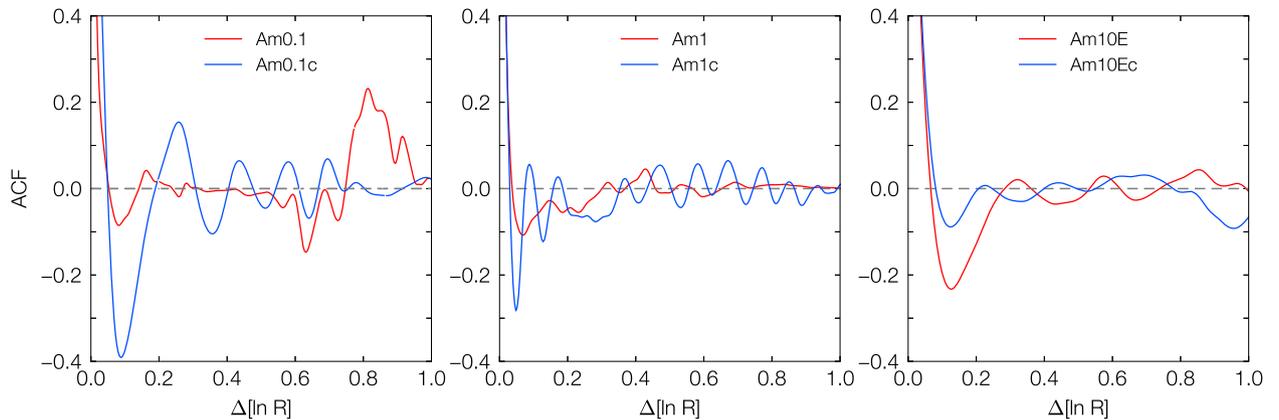}
\caption{The spatial auto-correlation function as a function of $\Delta \ln R$ (Equation \ref{eq:acf}). Blue curves denote ACF for simulations models with instantaneous cooling $\tau=0$. Red curves denote models with $\tau=1$. From left to right, the three panels present models with $Am=0.1$, $Am=1$, and $Am=10$ in Table \ref{table:runs}, respectively. 
}
\label{fig:acf}
\end{figure*}

\begin{figure*}
\centering
\includegraphics[width=0.95\textwidth]{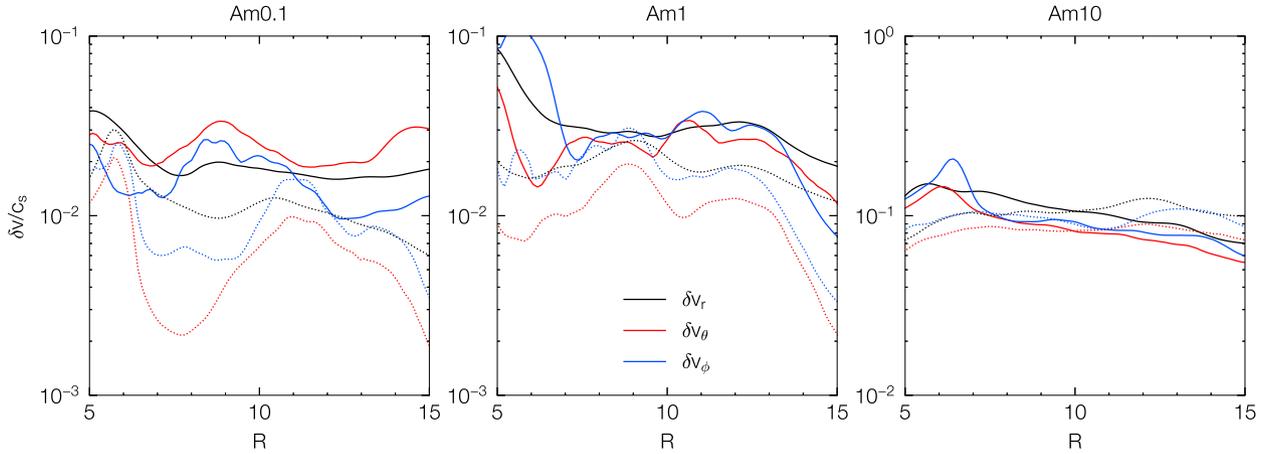}
\caption{
Three components of density-weighted $\delta v$ normalized by the local sound speed $c_s$ as a function of radius $R$. Solid curves denote models with instantaneous cooling $\tau=0$. Dotted curves denote models with $\tau=1$.
}
\label{fig:dv}
\end{figure*}

\begin{figure}
\centering
\includegraphics[width=0.5\textwidth]{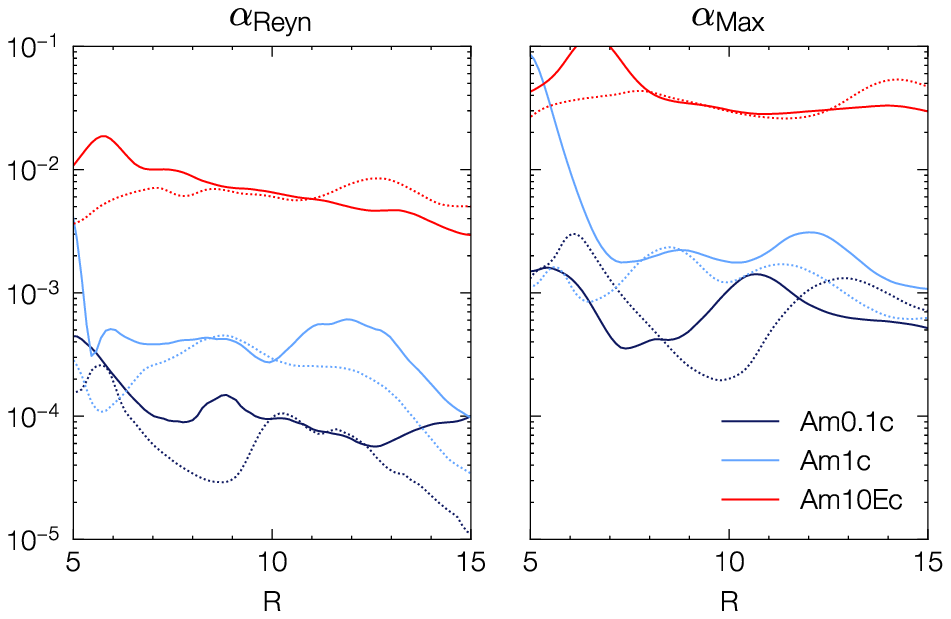}
\caption{
The Shakura-Sunyaev $\alpha$ parameter from Reynolds (left) and Maxwell (right) stress as a function of $R$ for Models \texttt{Am0.1c, Am1c, Am10Ec} (solid) and Models \texttt{Am0.1, Am1, Am10E} (dotted).
}
\label{fig:alpha}
\end{figure}

\subsection{The dominance between VSI and MRI}\label{sec:3.2}

Commonly reported in global hydrodynamic and non-ideal MHD simulations, the non-linear saturated state of the VSI is characterized by large-scale coherent vertical oscillations (e.g. \citealp{nelson_etal13,cb20}), which is inherited from its linear corrugation modes \citep{nelson_etal13,bl15}. Figure \ref{fig:vz} shows snapshots of azimuthally averaged vertical velocities $v_z$ at $t=2600P_0$ for different models. The top panels show simulation models with instantaneous cooling $\tau=0$. For comparison purposes, the bottom panels show the corresponding models from \citetalias{cb21} with the same $Am$ but different cooling timescale $\tau=1$.  
Comparing each column in Figure \ref{fig:vz}, it is clearly seen that the VSI has developed for model \texttt{Am0.1c} and \texttt{Am1c}, while for model \texttt{Am10Ec} the coherent vertical motions do not appear, and the MRI turbulence dominates. 

To quantify the effect of the VSI in $\tau=0$ simulations, we present  {\it spatial} auto-correlation function (ACF) for vertical velocities in Figure \ref{eq:acf}:
\begin{equation}
    \mathrm{ACF}(\Delta\ln R) = \frac{\langle V_z(R) V_z(R+\Delta R) \rangle} {\langle V_z^2 \rangle}.    
    \label{eq:acf}
\end{equation}
As our simulations are scale free, we adopt a normalized vertical velocity $V_z\equiv v_z/c_s$ at each radius, and present ACF in natural logarithm of $R$. The angle brackets $\langle\cdot \rangle$ denote the spatial averaging for a box spans over $[5,15]\times[-2H, 2H]\times[0,\pi/4]$ in $r\times\theta\times\phi$, and $t\in [2000,2600]P_0$. Note that by definition, the ACFs obtained at $\Delta \ln R=0$ should be strictly equal to unity.

In Figure \ref{eq:acf}, the left panel presents models with $Am=0.1$ (Table \ref{table:runs}). The ACF of model \texttt{Am0.1c} is clearly different from model \texttt{Am0.1}. Model \texttt{Am0.1c} has an alternating positive and negative pattern with strong correlation/anti-correlation amplitudes of $|\mathrm{ACF}|\sim0.1-0.2$. The characteristic length-scale of this variation is $\Delta \ln R \sim0.25$, equivalent to $\Delta R\sim 2.5H$. This lengthscale should correspond to the VSI corrugation modes wavelength. Indeed, it is roughly consistent with the wavelength seen in top-left panel of Figure \ref{fig:vz}. For model \texttt{Am0.1}, the oscillatory pattern is not seen, corresponding to the absence of the VSI.

The right panel of Figure \ref{fig:acf} shows models with $Am=10$. The ACFs have similar patterns for both models, where the amplitudes of ACFs first drop significantly and overshoot becoming negative, then almost vanishing for $\Delta \ln R\gtrsim0.2-0.3$. The similarity shared by the two curves in Figure \ref{fig:vz}, together with the lack of oscillatory pattern in ACF, suggest the prevalence of the MRI and the absence of the VSI in $Am=10$ models. 

The middle panel of Figure \ref{fig:acf} shows ACFs for $Am=1$ models. As in model \texttt{Am0.1c}, the ACF of model \texttt{Am1c} possesses similar alternating positive and negative patterns, though with lower amplitudes of $|\mathrm{ACF}|\sim0.05$. The characteristic length-scale is shorter with $\Delta \ln R \sim 0.1$, or equivalently $\Delta R\sim H$, consistent with the wavelength seen in the top-middle panel of Figure \ref{fig:vz}. We interpret such features as the competition between the VSI and the MRI turbulence, where larger-scale vertical oscillations of the VSI being partially disrupted by the MRI, leaving lower-amplitude, smaller radial lengthscale oscillations. In other words, our results imply that the VSI and the MRI coexist in model \texttt{Am1c}.

\begin{figure*}
\centering
\includegraphics[width=1\textwidth]{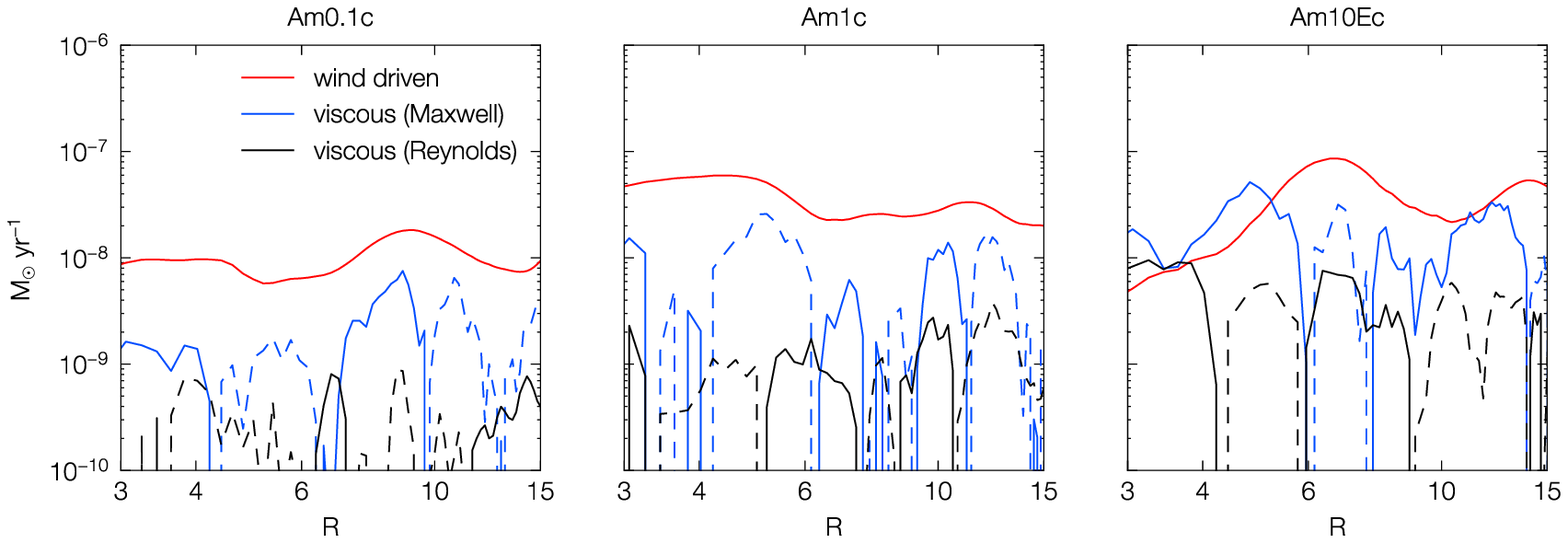}
\caption{
Mass accretion rate in $M_{\odot}\ {\rm yr}^{-1}$ of model $\texttt{Am0.1c}$, $\texttt{Am1c}$, and $\texttt{Am10Ec}$. Accretion driven by MHD winds (red), Maxwell stress (blue), and Reynolds stress (black) are shown, computed by equations (18), (19), and (20) in \citet{cb20}, respectively. Solid and dashed curves denote positive and negative values.
Similar plots for simulation models with $\tau=1$ can be found in Figure 8 and Figure 13 of \citetalias{cb21}.}
\label{fig:AM}
\end{figure*}

\subsection{Turbulence levels and stresses}\label{sec:3.3}

We quantify disk turbulence strengths by computing the root mean square of velocity fluctuations,
\begin{equation}
\delta v =\Bigg[\frac{1}{n}\sum\limits_{i=1}^n(v_i-\langle v \rangle)^2\Bigg]^{1/2},
\end{equation}
where $\langle v \rangle$ denotes azimuthal averaging.
Figure \ref{fig:dv} shows the radial profiles of three components of density-weighted turbulent velocities $\delta v/c_s$, normalized by local sound speed. Spatial averaging is performed over $\Delta R \in [0.95R, 1.05R]$ and $\Delta z \in [-3H, 3H]$, and temporal averaging over the last $400P_0$. For models \texttt{Am0.1c} and \texttt{Am1c}, the presence of the VSI enhances the turbulent velocities to $\delta v/c_s\sim0.02-0.04$, comparing to their $\tau=1$ counterparts with the absence of the VSI. In model \texttt{Am0.1c}, the large amplitudes in meridional turbulent velocities $\delta v_\theta$ is associated with VSI coherent vertical oscillations. In model \texttt{Am10Ec}, the turbulent velocities reach $\delta v/c_s\sim 0.1$, comparable to model \texttt{Am10E}, as both models are dominated by the MRI turbulence. 

Figure \ref{fig:alpha} shows radial profiles of the Shakura-Sunyaev $\alpha$ parameter for Reynolds and Maxwell stress. The Reynolds stress is found to be $\alpha\sim1\times10^{-4}, \alpha\sim5\times10^{-4}$, and $\alpha\sim1\times10^{-2}$ in the domain $R\in[5,15]$ for models \texttt{Am0.1c}, \texttt{Am1c}, and \texttt{Am10Ec}, respectively. The Maxwell stresses are $\alpha\sim1\times10^{-3}, \alpha\sim3\times10^{-3}$, and $\alpha\sim3\times10^{-2}$, respectively. For reference, hydrodynamic VSI simulations incorporating different physics and spatial dimensions suggest a Reynolds stress of $\alpha\sim 10^{-4}-10^{-3}$ \citep{nelson_etal13,sk14,manger+20,pk21}. Our 3D non-ideal MHD simulation results reside in this range for VSI dominated models.

\subsection{Angular momentum transport}\label{sec:3.5}

To illustrate angular momentum transport in our simulations, Figure \ref{fig:AM} shows the mass accretion rate driven by MHD winds and by turbulent stresses in units of mass of the sun per year. The mass accretion rate in code units is converted to physical units by equation (25) of \citetalias{cb21},
\begin{equation}
\begin{split}
    &\dot{M}_{\rm phys} \approx 1.54\times10^{-8} M_{\odot}\ {\rm yr}^{-1} \bigg(\frac{\dot{M}_{\rm code}}{10^{-4}}\bigg)\bigg(\frac{M_\star }{M_\odot}\bigg)^{1/2}\\
    &\quad \times\bigg(\frac{R_{\rm phys}}{\rm 30AU}\bigg)^{1/2}\bigg(\frac{R_{\rm code}}{R_0}\bigg)^{q_D-3/2}\bigg(\frac{\Sigma}{10{\rm g}\ {\rm cm}^{-2}}\bigg) \ .
\label{eq:mdot}     
\end{split}
\end{equation}
The above expression is interpreted as follows. If we consider at the radius $R=1$ in code units ($R_{\rm code}=1R_0$) to correspond to a physical radius of $30$ AU, where the surface density is $10$ g cm$^{-2}$, then the physical accretion rate can be directly read off from equation \eqref{eq:mdot} with $\dot{M}_{\rm code}$ obtained from simulation, and is plotted in Figure \ref{fig:AM}.

In Figure \ref{fig:AM}, all three models possess accretion rates on the order of $10^{-8}-10^{-7} \ M_{\odot}\ {\rm yr}^{-1}$, consistent with observed mass accretion rates \citep{hartmann_etal98,hh08}. Mass accretion is predominantly mediated by MHD winds for all three models, as being also witnessed in \citetalias{cb21}. For models $\texttt{Am0.1c}$ and $\texttt{Am1c}$, viscous driven accretion by MRI and VSI turbulence makes minor contributions. For model $\texttt{Am10Ec}$, Maxwell stress contributes nearly equally to MHD winds due to the vigorous MRI turbulence. Accretion driven by the Reynolds stress is small compared to that of Maxwell stress and MHD winds in all models.  

\subsection{Magnetic flux concentration and ring formation}\label{sec:3.4}

\begin{figure*}
\centering
\includegraphics[width=0.95\textwidth]{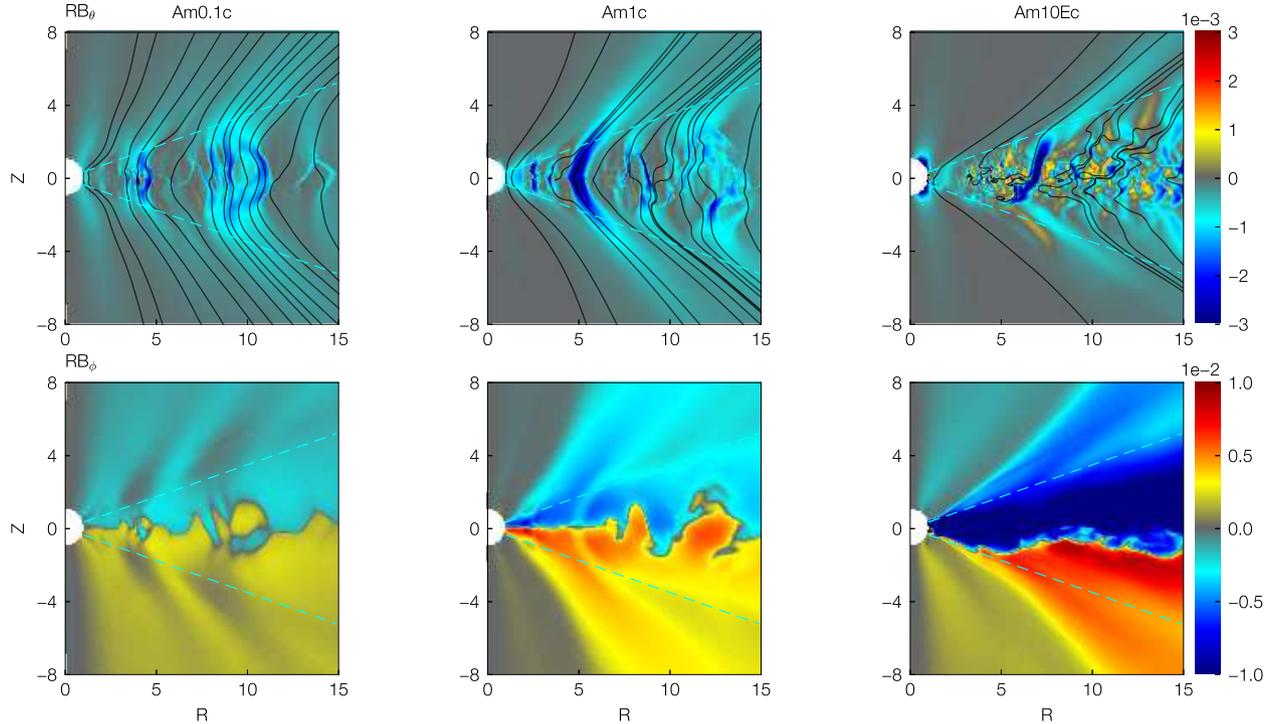}
\caption{
Snapshots of azimuthally averaged magnetic field for model $\texttt{Am0.1c}$ (left), model $\texttt{Am1c}$ (middle), and Model $\texttt{Am10Ec}$ (right) at $t = 3000P_0$. Top panels: snapshots of scaled meridional magnetic field $RB_\theta$. Bottom panels: snapshots of scaled toroidal magnetic field $RB_\phi$. Overlaid black curves are equally spaced contours of poloidal magnetic flux. Dashed lines mark an opening angle of $0.35$ above and below the midplane, which correspond to the transition from disk zone to wind zone.
}
\label{fig:B}
\end{figure*}

\citetalias{cb21} demonstrates the presence of spontaneous magnetic flux concentration in strong ambipolar diffusion dominated ($Am\leq3$) , weakly MRI turbulent disks. On the other hand, weak ambipolar diffusion ($Am=10$) permits rigorous MRI turbulence that disrupts the concentrated flux sheets. Here, we show that the flux concentration can sustain in strong ambipolar diffusion dominated, VSI turbulent disks. For model \texttt{Am0.1c} and \texttt{Am1c}, Figure \ref{fig:B} (top panels) and Figure \ref{fig:st} illustrate that the poloidal magnetic fields assemble into magnetic flux sheets at different radial locations\footnote{For model \texttt{Am10Ec}, there is a thick magnetic flux sheet that appears at a radius $R\sim5-7$ in the upper right panel of Figure \ref{fig:B}, which seems inevitable in the $Am=10, \tau=0$ simulations as we experimented. This might be caused by the toroidal field of negative sign that overwhelms over the disk column for $R\lesssim5$ seen in the lower right panel of Figure \ref{fig:B}, squeezing the poloidal field into a slanted configuration.}. The flux sheets are nearly stationary for model \texttt{Am0.1c} and slowly propagating radially outwards for model \texttt{Am1c}. 

The inhomogeneous distribution of poloidal magnetic fields leads to variations in surface densities. Figure \ref{fig:sigma} shows the radial profiles of surface density and vertical magnetic field strength for models \texttt{Am0.1c} and \texttt{Am1c}.
%Local surface density maxima and minima are developed.
%To quantify widths and depths, we fit a Gaussian of functional form $A_0+A\cdot e^{\frac{-(x-x_0)^2}{2\sigma^2}}$ to each of the bumps; $\sigma$ is the standard deviation and $x_0$ is the location of the bump center. We define the width in units of gas scale height to be $\sigma/(hx_0)$ and the variation in depth to be $A/A_0$. 
%{\bf The fittings give widths of $\sim1H, 1.25H$ and variations in depths of $\sim100\%, 170\%$ for the two bumps located at $R=4$ and $R=8$ in model \texttt{Am0.1c}. The widths are $\sim1.3H,0.7H,0.7H$ and variations in depths are $\sim 200\%,40\%,25\%$ for bumps at $R=4$, $R=7$, and $R=10$ in model \texttt{Am1c}. We note that there is a deep gap carved in gas surface density at $R=5$ in model \texttt{Am1c}. It is not a unique feature for this model as it also appeared in \texttt{B3} and \texttt{Am3E} in \citetalias{cb21} (see their Figure 5), and likely caused by the stochastic nautre of the magnetic flux concentration phenomenon.}
Overall, the radial variations in surface density profile is comparable to those in \citetalias{cb21}. Gas density bumps are developed at radii $R=4,8$ in model \texttt{Am0.1c} and radii $R=4,7,10$ in model \texttt{Am1c}. They are separated by several gas scale heights ($\gtrsim5H$). The surface density variations, measured by the peak-to-trough ratio minus one, in the bumps are typically modest (of order unity), except for a deep gap at $R=5$ in model \texttt{Am1c}. It is not a unique feature for this model but also appeared in \texttt{B3} and \texttt{Am3E} in \citetalias{cb21} (see their Figure 5), and likely caused by the stochastic nature of the magnetic flux concentration phenomenon.

%$\sim1H,1.3H$, and variations in bump depths of $\sim60\%$ and $\sim16\%, 60\%$ for models \texttt{Am0.1c} and \texttt{Am1c}, respectively.} These results are comparable with $\tau=1$ models in \citetalias{cb21}. 

\subsubsection{Mechanism: advection or ambipolar diffusion?}\label{sec:3.4.1}

Various physical mechanisms have been proposed for magnetic flux concentration in non-ideal MHD simulations of PPDs. We point the reader to a summary in section 6.2 of \citetalias{cb21}. As these mechanisms involve distinct physical processes and are still under debate, here we do not seek to verify them. Instead, we aim to distinguish the role between advection and (ambipolar) diffusion of the magnetic field in flux concentration phenomenon in our simulations.   

The rate of change of magnetic flux is described by the Faraday's Law. In its integral form, the rate of change of poloidal magnetic flux is related to the toroidal electric field \citep[e.g.,][]{lubow94,bs17},
\begin{equation}
    \frac{1}{c}\dv{\Phi}{t} = -2\pi R \langle E_\phi\rangle,
\label{eq:F}
\end{equation}
where the poloidal magnetic flux is 
\begin{equation}
\Phi(r,\theta)= \int_0^\theta \ \langle B_r(r,\theta)\rangle r^2 \sin\theta \ d\theta\,
\end{equation}
and the toroidal electric field is
\begin{equation}
    E_\phi = [\vb{v}\times\vb{B}]_\phi-\frac{4\pi}{c}{\eta_{\rm A}\rm J_{AD,\phi}}.
\label{eq:E}
\end{equation}
On the right hand side, the first term corresponds to the advection of magnetic field $\vb{B}$ with the fluid at velocity $\vb{v}$, and the second term arises from the diffusion of magnetic filed through the fluid due to ambipolar diffusion; $\eta_{\rm A}$ is the ambipolar diffusivity, $\vb{J}_{\rm AD} = -(\vb{J} \times \vb{b}) \times \vb{b}$ is the effective current density associated with ambipolar diffusion, where $\vb{J}=c\curl\vb{B}/4\pi$ is the current density, and $\vb{b}=\vb{B}/B$ is the unit vector of magnetic field. The angle brackets $\langle\cdot\rangle$ denote azimuthal averaging, and above equations are written in Gaussian units. 

\begin{figure}
\centering
\includegraphics[width=0.5\textwidth]{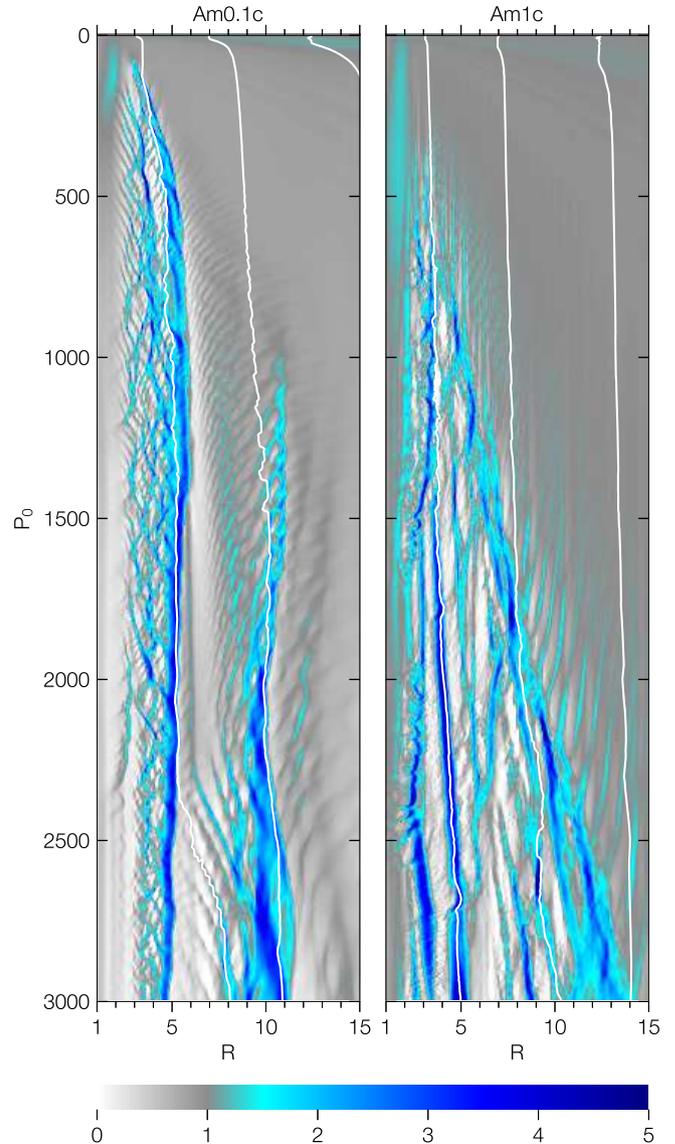}
\caption{
Space-time diagram of azimuthally and vertically ($\pm1 H$) averaged midplane $B_z/B_{z,t = 0}$ for Model $\texttt{Am0.1c}$ (left) and Model $\texttt{Am1c}$ (right). Overlaid white vertical curves delineate linearly equally spaced contour lines of poloidal magnetic flux $\Phi$. 
Similar plot for simulation models with $\tau=1$ can be found in Figure 4 of \citetalias{cb21}.
}
\label{fig:st}
\end{figure}

\begin{figure}
\centering
\includegraphics[width=0.5\textwidth]{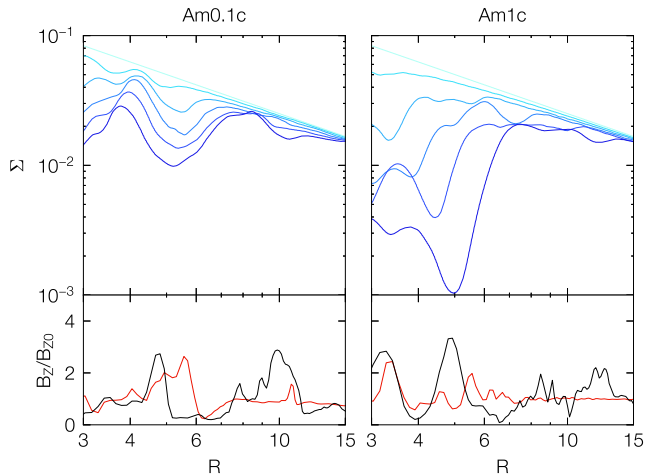}
\caption{
Top panels: surface densities $\Sigma$ as a function of radius at $t=0,600,1200,1800,2400,3000P_0$ from light to dark colors. Bottom panels: vertical magnetic fields normalized by initial values $B_z/B_{z,t=0}$ as a function of radius at $t=1200P_0$ (red) and $t=3000P_0$ (black).
Similar plot for simulation models with $\tau=1$ can be found in Figure 5 of \citetalias{cb21}.
}
\label{fig:sigma}
\end{figure}

\begin{figure*}
\includegraphics[width=0.95\textwidth]{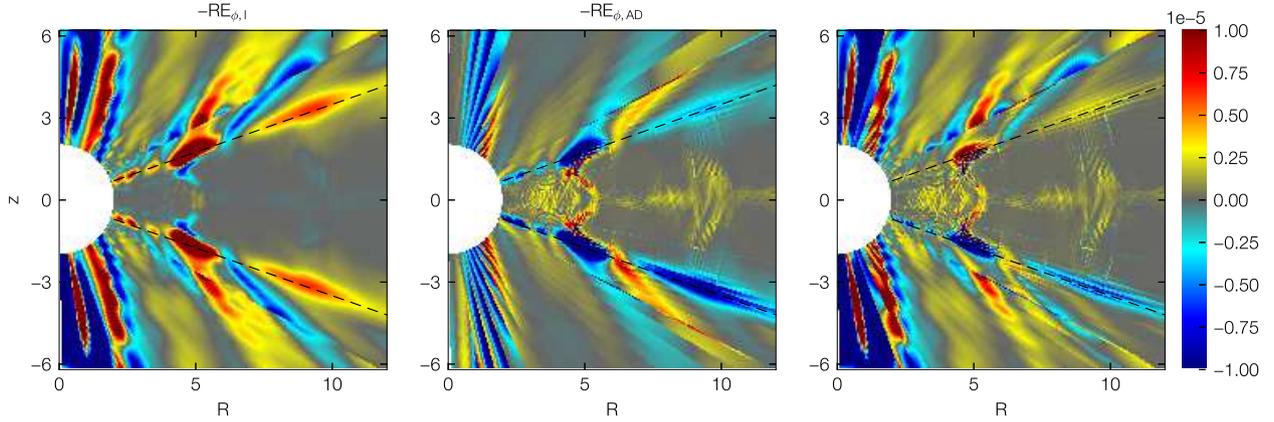}
\caption{The advection component (left), ambipolar diffusion component (middle), and total (right) of $-RE_\phi$ for model $\texttt{Am0.1c}$ at $t=2000P_0$, averaged over $\pm200P_0$.
}
\label{fig:flux}
\end{figure*}

Taking model $\texttt{Am0.1c}$ as an example, Figure \ref{fig:flux} shows the rate of change of poloidal magnetic flux, represented by $-RE_\phi$, at a evolutionary time of $t=2000P_0$ averaged over $\pm200P_0$. Over this period, magnetic flux continues to build up at two radial intervals, $R\sim3-5$ and $R\sim8-11$, as seen in Figure \ref{fig:st}. Flux accumulation (depletion) requires a positive (negative) radial gradient in $-RE_\phi$. Indeed, the right panel of Figure \ref{fig:flux} reveals prominent features in total $-RE_\phi$ at those two radial intervals, which reflect the local increase or decrease in magnetic flux, for which local increase corresponds to magnetic flux concentration. The similarity between the middle and right panels indicates that ambipolar diffusion is the major contributor to the accumulation of poloidal magnetic flux, whereas contribution from advection is minor as evident from the left panel. This result is also valid for model $\texttt{Am1c}$ and for models in \citetalias{cb21} with $Am \leq 3$.

\begin{figure*}
\includegraphics[width=0.95\textwidth]{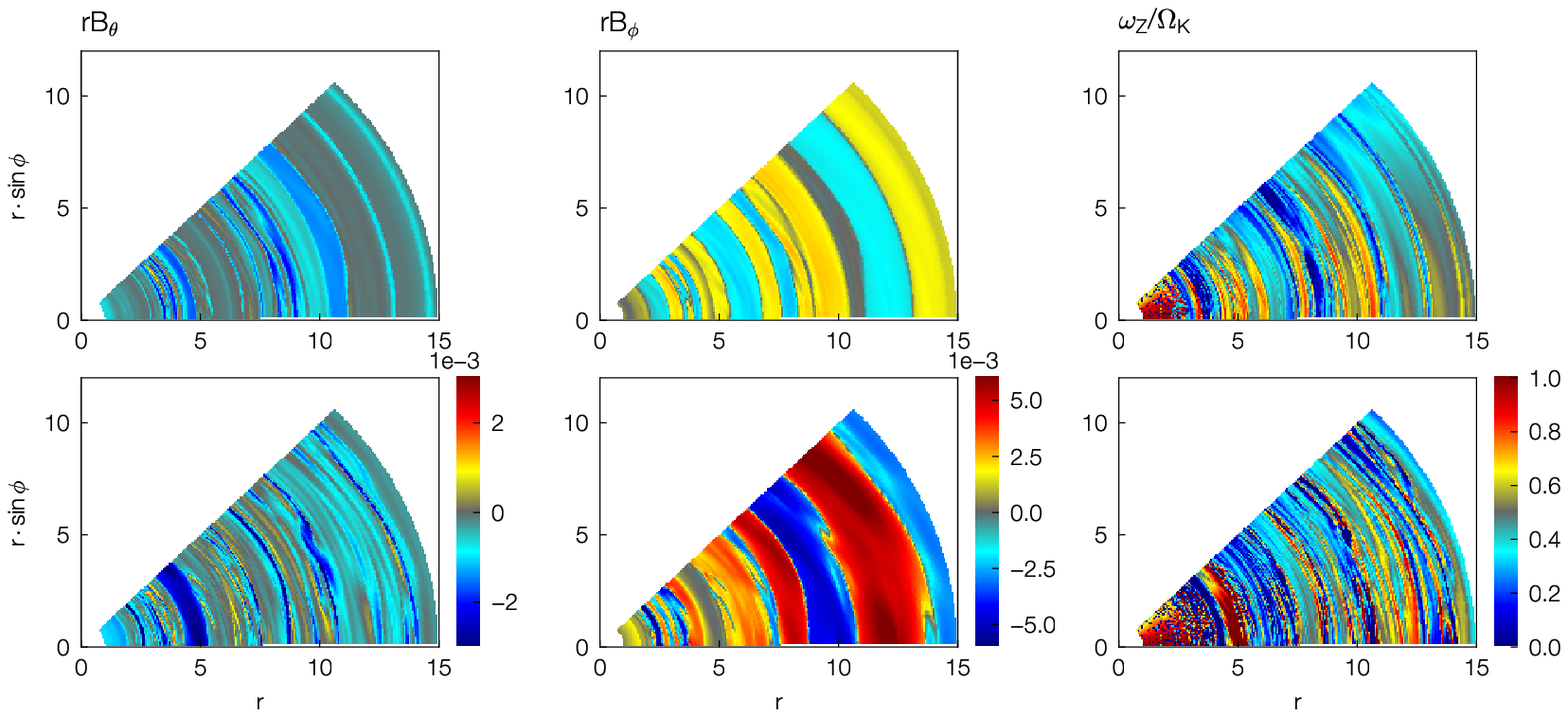}
\caption{
Snapshots of scaled meridional magnetic field $rB_\theta$ (left), azimuthal magnetic field $B_\phi r$ (middle), and vorticity $\omega_z/\OmK$ (right) at the midplane in the $R-\phi$ plane, for Model $\texttt{Am0.1c}$ (top) and Model $\texttt{Am1c}$ (bottom) at $t=3000P_0$. Similar plot can be found in Figure 10 of \citetalias{cb21}.
}
\label{fig:Rphi}
\end{figure*}

\subsection{Azimuthal vortices}\label{sec:3.6}

In 3D hydrodynamic simulations of the VSI, vortices in the $R-\phi$ plane are commonly reported \citep[e.g.,][]{richard+16,manger+20}. On the other hand, surface density variations induced by the magnetic flux concentration could trigger Rossby wave instability (RWI) and subsequently produce vortices. 
However, seen in the right column of Figure \ref{fig:Rphi}, no vortices are found in all of our simulation models, regardless of the value of $Am$, despite that the azimuthal resolution is sufficiently fine to resolve VSI and RWI induced vortices \citep[e.g.,][]{manger+20,li+01}. We suspect the absence of vortices might be caused by 1) the limited azimuthal domain which requires high surface density variations to trigger RWI \citep{ono+16}. Taking case (iii) in \citet{ono+16} as an example, which is closest to our simulation model, their Figure 7 demonstrates that to excite $m=7$ or higher-order RWI modes, the surface density variation shall satisfy peak-to-trough ratio much larger than unity, exceeding that found for model $\texttt{Am0.1c}$ in section \ref{sec:3.4}; 2) the presence of magnetic field may prevent the rolling up of vortex sheets by magnetic tension \citep{lyra11}. Future studies should address this issue.

\section{Conclusions and Discussion} \label{sec:cd}

This paper extends the previous work of \citetalias{cb21} to examine the
interplay between VSI and MRI in ambipolar diffusion dominated outer PPDs, via global 3D non-ideal MHD simulations. Given the range of ambipolar Els\"{a}sser numbers explored, our main findings are:
\begin{itemize}
\item The VSI turbulence dominates over the MRI when ambipolar diffusion is strong ($Am=0.1$), clearly showing VSI coherent vertical oscillations, enhancing the level of turbulence to $\delta v/c_s\sim0.02-0.03$. The Shakura-Sunyaev $\alpha$ parameter reaches $\alpha\sim10^{-4}$ and $\alpha\sim10^{-3}$ for Reynolds and Maxwell stresses, respectively.
\item The VSI and MRI co-exist for $Am=1$, with VSI-like though less coherent vertical oscillations. The level of turbulence reaches $\delta v/c_s\sim0.03$, and $\alpha\sim5\times10^{-4}$ and $\alpha\sim3\times10^{-3}$ for Reynolds and Maxwell stress, respectively.
\item The VSI is overwhelmed by the MRI when ambipolar diffusion is weak ($Am=10$), reaching high level of turbulence with $\delta v/c_s\sim0.1$. MRI turbulence results in strong Reynolds and Maxwell stress, with $\alpha\sim10^{-2}$ and $\alpha\sim3\times10^{-2}$, respectively.
\end{itemize}
Angular momentum transport is dominated by magnetized disk winds in all models, where viscous accretion via turbulence plays a moderate role in most cases. 
Formation of annular substructures by spontaneous magnetic flux concentration remains robust in the VSI turbulent disks (model $\texttt{Am0.1c}$ and $\texttt{Am1c}$). The magnetic flux concentration is primarily attributed to the ambipolar diffusion.

The dominance between the VSI and the MRI with the presence of magnetic fields has been studied in linear theory. In the ideal MHD limit, the MRI linear modes grow faster than the VSI by a factor of $\sim R/H$ \citep{lp18}, while non-ideal MHD effects can rescue the VSI and dampen the MRI \citep{cl21,LK22}. Our simulation results show consistency with the linear theory for small ($Am=0.1$) and large ($Am=10$) ambipolar Els\"{a}sser numbers. For the intermediate case of $Am=1$, our results for the first time demonstrate the co-existance of the VSI and the MRI turbulence in ambipolar diffusion dominated outer disks. 

We employ a disk equilibrium temperature gradient to be $T\propto r^{-q_T}$, with power-law index $q_T=1$. In a realist disk, this power law is expected to be shallower $q_T\approx1/2$ \citep{cg97}, which will result in weaker VSI turbulent velocities $\delta v$ and smaller VSI-induced $\alpha$ values \citep{nelson_etal13,manger+21}. Moreover, instantaneous cooling ($\tau=0$) is employed in all models of this work. The cooling timescale in a realistic disk depends on two processes: the thermal coupling between the gas and dust, and the optical depth \citep{bae+21}. Taken these processes into account, it is found that greater than $\sim10$ AU, the fast cooling timescales could permit the VSI growth \citep{pk21}. But these timescales are also finite, which will result in weaker $\delta v$ and smaller $\alpha$ compared to the instantaneous cooling adopted in this work \citep{nelson_etal13,manger+21}.

The coherent vertical oscillations driven by corrugation modes is a unique feature of the VSI, which can help differentiate itself from other dynamical processes in PPDs \citep[e.g.,][]{flock_etal20,BA+21,Blanco21}. Although in model \texttt{Am0.1c} the dominant radial lengthscale of the corrugation modes is relatively large, in model \texttt{Am1c} the interplay between VSI and MRI results in less coherent oscillations with narrower radial lengthscales of $\sim H$, making it difficult to be identified and spatially resolved. Further complexity can arise from the inertial wave parametric instability that attacks the VSI corrugation modes, producing even smaller-scale turbulence in disks \citep{cl22,scl22}.  

This work along with \citetalias{cb21} demonstrates the robustness of magnetic flux concentration phenomenon and annular substructure formation in ambipolar diffusion dominated regions. To confront observations with theory, dust particles shall be incorporated in our simulations to study its dynamics under the influence of magnetic flux concentration. In MRI-turbulent local shearing box simulations, \citet{xb21} found that magnetic flux concentration leads to dust clumping at pressure maxima. It would be intriguing to test this finding in 3D global simulations, possibly with the presence of VSI turbulence. Moreover, the VSI and MRI may possess non-Kolmogorov energy spectra, and it could modify the grain collisional velocities calculated from classic Kolmogorov turbulence \citep{oc07,gong+20,gong+21}.

Future study should focus on relaxing the limited azimuthal domain, in particular to examine the existence and evolution of vortices. It would also be useful to further incorporate dust particles in our global non-ideal MHD simulations and to explore dust trapping in pressure bumps induced by the magnetic flux concentration. Lastly, it is necessary to employ sufficient fine resolution to resolve the parametric instability that attacks the VSI corrugation modes and drives small-scale inertial wave turbulent cascade.

\section*{Acknowledgements}

We thank our referee, Zhaohuan Zhu, for a set of useful comments that improved the presentation of this work.
We thank William B\'{e}thune, Cornelis Dullemond, Munan Gong, Anders Johansen, Michiel Lambrechts, Henrik Latter, Marius Lehmann, Geffroy Lesur, Min-Kai Lin, Sebastian Marino, Sijme-Jan Paardekooper, Richard Teague, and Andrew Youdin for the useful discussions. CC acknowledges the support from AFD group at Department of Applied Mathematics and Theoretical Physics, University of Cambridge, and STFC grant ST/T00049X/1. XNB acknowledges the National Key R\&D Program of China No. 2019YFA0405100. Numerical simulations are conducted on the Orion cluster at Department of Astronomy, Tsinghua University, and on TianHe-1 (A) at National Supercomputer Center in Tianjin, China.

\section*{Data Availability}

The data underlying this article will be shared on reasonable request to the corresponding author.

%%%%%%%%%%%%%%%%% APPENDICES %%%%%%%%%%%%%%%%%%%%%

%\appendix
%\section{}\label{sec:}

%%%%%%%%%%%%%%%%%%%% REFERENCES %%%%%%%%%%%%%%%%%%

\bibliographystyle{mnras}
\bibliography{disk} 

%%%%%%%%%%%%%%%%%%%%%%%%%%%%%%%%%%%%%%%%%%%%%%%%%%
\bsp	
\label{lastpage}
\end{document}